\documentclass[conference,dvipsnames]{IEEEtran}

\usepackage{tikz}
\usetikzlibrary{arrows.meta,arrows,automata,positioning}
\usepackage{amsmath}

\usepackage{booktabs}
\usepackage{microtype}
\usepackage{graphics}
\usepackage{balance}
\usepackage[htt]{hyphenat}

\usepackage[hidelinks]{hyperref}
\graphicspath{{figures/}}
\usepackage{multirow}
\usepackage{caption}
\usepackage{subcaption}
\usepackage[english]{babel}

\usepackage{comment}
\usepackage{listings}
\usepackage{todonotes}

\newcommand{\exhaustive}[0]{dense}
\newcommand{\Exhaustive}[0]{Dense}

\usepackage[backend=bibtex,style=ieee,mincitenames=1,maxcitenames=1,doi=false,isbn=false,url=false,eprint=false]{biblatex}
\DefineBibliographyStrings{english}{%
  andothers = {et al\adddot}
}

\usepackage[inline]{enumitem}
\setlist[itemize]{noitemsep}

\newcommand{\mad}{\textsc{Mad}}
\newcommand{\rh}{RowHammer}

\newcommand{\buddyL}[1]{$\phi$}
\newcommand{\buddyR}[1]{$\psi$}

\begin{document}

\title{\mad{}: Memory Allocation meets Software Diversity}

\author{
	\IEEEauthorblockN{Manuel Wiesinger}
	\IEEEauthorblockA{Vrije Universiteit Amsterdam\\
	m.wiesinger@vu.nl}
	\and
	\IEEEauthorblockN{Daniel Dorfmeister}
	\IEEEauthorblockA{Software Competence Center Hagenberg\\
	daniel.dorfmeister@scch.at}
	\and
	\IEEEauthorblockN{Stefan Brunthaler}
	\IEEEauthorblockA{$\mu$CSRL -- Research Institute CODE\\
	Universit{\"a}t der Bundeswehr M{\"u}nchen\\
	brunthaler@unibw.de}
}


\maketitle

\begin{abstract}
  Vulnerabilities emanating from DRAM errors pose a vexing problem that remains, as of yet, unsolved and elusive but cannot be ignored.
  Prior defenses focused on specific details of early \rh{} attacks and fail to generalize with the generalizations of recent \rh{} attacks.
  Even worse, it is presently not clear that techniques from prior defenses will be able to cope with these generalizations or if an entirely new approach is required.
  Although still work-in-progress, we have identified a new approach that combines memory allocation with principles underlying software diversity and shows promising early results.

  At first glance, software diversity seems to be an unlikely contender, since it faces seemingly insurmountable obstacles, primarily the lack of sufficient entropy in memory subsystems.
  Our system---called \mad{}, short for memory allocation diversity---leverages two novel, complementary spatial diversification techniques to overcome this entropy obstacle.
  Entropy aside, \mad{} offers ease-of-implementation, negligible performance impact, and is both hardware and software agnostic.

  From a security perspective, \mad{}'s goal is to deter \rh{} attacks by delaying them to the maximum extent possible.
  Such a delay opens the door for a variety of additional responses, e.g., proactive rebooting, or complementary in-depth analysis of ongoing attacks that would be too slow for an always-on defense.
\end{abstract}


\section{Diversity meets \rh{}}
\label{s:motivation}

The \rh{} vulnerability, published in \citeyear{rowhammer}~\cite{rowhammer}, has taken the world by surprise and dealt a severe blow to one of the core tenets of operating systems, namely the integrity of their internal data structures.
Without actually accessing internal OS data structures (e.g., page tables), \rh{} attacks showed how to manipulate these data by row hammering adjacent memory rows.

Recent results demonstrate that \rh{} is still possible on DDR4 DRAM devices that use the hardware defense target row refresh (TRR)~\cite{Frigo2020,Ridder2021}, and that ECC memory---contrary to initial thoughts---provides no safe haven~\cite{eccploit}.
What follows from these recent developments is that much of the prior approaches to prevent \rh{} attacks provide inadequate protection.

Two important generalizations over the state-of-the-art from the early period of \rh{} attacks are as follows:
(i)~spatial co-location helps but is not an indispensable prerequisite (single/double-sided \rh{} extended to many-sided \rh{}); and
(ii) error correction puts constraints on which susceptible bit flips an attack may use.
In an upcoming paper, we also see that exclusive protection in operating systems is inadequate, as JavaScript continues to offer sufficient attack surface~\cite{rowhammerjs,Ridder2021}.

A closer investigation of \rh{} attacks shows that they have one thing in common: they require a form of memory massaging~\cite{drammer, flip_feng_shui, another_flip_in_the_wall}.
This memory massaging is required to obtain a vulnerable configuration, i.e., a configuration that is amenable to adversarial control.
The vulnerable configuration consists of (i) a memory allocation that the attacker needs for \rh{}, i.e., control of the rows that, if subjected to \rh{}, trigger a bit flip in some other target location, as well as (ii) forcing a third party to put target data into that target location.
This third party often is a memory allocator.

Three implications arise from these observations.
First, the adversary requires a reconnaissance phase to identify the vulnerable configuration, i.e., the specific rows he needs to control and which rows hold flippable bits suitable for target data manipulations.
Second, the adversary needs to acquire control of the vulnerable configuration.
Third, the adversary needs a way to coerce and coopt the third party, e.g., through predictability of a memory allocator.
Note that the first and second stages may be combined.

We have identified two different strategies of abusing predictable memory allocators:
(i) \emph{\exhaustive{}-allocation massaging}, e.g., Flip-Feng Shui~\cite{flip_feng_shui} or memory waylaying~\cite{another_flip_in_the_wall}, and
(ii) \emph{sparse-allocation massaging}, e.g., Phys-Feng Shui~\cite{drammer} or memory chasing~\cite{another_flip_in_the_wall}.
\Exhaustive{} means that the adversary allocates and holds all memory, whereas in sparse allocation, she tries to allocate all, but hold on to as little memory as necessary.

Memory spraying techniques, exemplified by Seaborn and Dullien's Rowhammer attack~\cite{project_zero_rowhammer}, combine aspects of these two allocation strategies.
Instead of allocating \emph{all} memory, which could trigger operating system intervetion, spraying uses an \exhaustive{} approach to allocate large partitions of memory, e.g., allocating a third of all available memory.
If spraying did not find a vulnerable configuration, then the memory partition will be released, and a new attempt will be made---effectively resembling a sparse-allocation.

Software Diversity belongs to the area of biologically-inspired software defenses, with the core principle to overcome negative effects of a monoculture.
In code-reuse attacks, adversaries enjoy large economies of scale through executable programs being identical across vast numbers of machines.
In \rh{} attacks, adversaries enjoy similar benefits through the predictability of operating systems, specifically memory allocation strategies and management of internal data structures.

The key difference between these attacks is their susceptibility and amenability to diversification transformations.
A diversifying compiler can, for example, randomize essentially all aspects of an executable, and since there is essentially no hard limitation for executable size, diversity remains effective.
Transplanting the core principles underlying diversity to the domain of memory allocation brings about an important challenge: the lack of entropy in memory allocation.
A memory allocator, on the other hand, has a hard limit, namely the fixed amount of physical memory present in a computer.
Consider, for example, a system with 16 GB of RAM, the memory allocator manages merely four million 4KB memory pages, thus severely limiting the applicability and effectiveness of traditional diversification techniques.

\mad{} combines two complementary novel, spatial diversification techniques that overcome the entropy obstacle and prolong both allocation strategies.
Prolonging \exhaustive{}-allocation massaging gives us the opportunity to maximize the likelihood of detecting such an attack.
Prolonging sparse-allocation massaging allows us to increase the time required for the attack to succeed, ideally such that performing the attack never succeeds.
Since memory spraying techniques combine aspects of both massaging techniques, \mad{} prolongs spraying, too.

Summing up, the contributions of this paper are as follows:
\begin{itemize}
	\item We introduce \emph{memory allocation diversity}, \mad{} for short, a method to diversify memory management.
	      At its core, \mad{} uses a diversified cache that manages the memory blocks obtained from an underlying memory manager.
  \item We illustrate two novel, spatial diversification techniques that combine to deter the memory massaging part of a \rh{} attack.
	\item We subjected the prototype to a variety of different experiments to evaluate its security.
      Our early results look promising, they indicate that \mad{} delays sparse-allocation memory massaging and offers leverage to detect \exhaustive{}-allocation massaging.
\end{itemize}

\section{Threat Model and Assumptions}
\label{ss:threat-model}

In our threat model, the adversary performs memory massaging as a by-product of the actual attack.
By exploiting \rh{}, e.g., an adversary may be interested in performing privilege escalation.
Alternatively, however, an attack may focus on altering information in a web browser by row hammering through JavaScript~\cite{rowhammerjs,Ridder2021}.
Since \mad{} itself generalizes to both domains---web browsers and operating systems---the corresponding threat model differs.
To this end, we focus on the operating system domain in this paper.

Our assumptions include the following:
\begin{itemize}
	\item The kernel is considered to be safe.
	      As a result, the attacker cannot modify kernel internals or tamper with \mad{}'s state or with the kernel's random number generator.
	\item The attacker can execute unprivileged code on a system.
	      From the perspective of \mad{}, it does not matter whether the attacker is working on a remote, virtual image, or on a local machine.
	\item The attacker cannot access the page map of the attack.
	      This assumption is not a strong requirement but serves as a simplification, as the information would help the attacker but is not sufficient to break \mad{}, as the attacker cannot manipulate the random number generator.
\end{itemize}

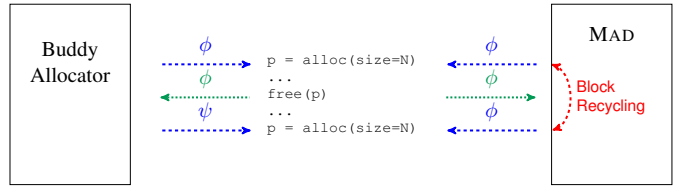
\begin{figure}[t!]
	\centering
	\scalebox{.4}{%
		\begin{tikzpicture}[>=stealth',shorten >=1pt,auto,node distance=2.8cm]

     \begin{scope}[]
       \node[scale=2.0,anchor=east] (srcCode) at (-5,0) {
            \begin{minipage}{2.9cm}
\begin{lstlisting}[frame=none]
  p = alloc(size=N)
  ...
  free(p)
  ...
  p = alloc(size=N)
\end{lstlisting}
            \end{minipage}%
        };
     \end{scope}

     \begin{scope}[shift={(.5,0)}]
       \draw[->,dashed,blue,line width=.66mm] (-2,1) to node[midway,above,scale=2.0]{$\phi$} (-5,1);
       \draw[->,dashed,blue,line width=.66mm] (-2,-1.2) to node[midway,above,scale=2.0]{$\phi$} (-5,-1.2);

       \draw[->,dotted,ForestGreen,line width=.66mm] (-5,-0.1) to node[midway,above,scale=2.0]{$\phi$} (-2,-0.1);
     \end{scope}

     \begin{scope}[shift={(-9,0)}]
       \draw[<-,dashed,blue,line width=.66mm] (-2,1) to node[midway,above,scale=2.0]{$\phi$} (-5,1);
       \draw[<-,dashed,blue,line width=.66mm] (-2,-1.2) to node[midway,above,scale=2.0]{$\psi$} (-5,-1.2);

       \draw[<-,dotted,ForestGreen,line width=.66mm] (-5,-0.1) to node[midway,above,scale=2.0]{$\phi$} (-2,-0.1);
     \end{scope}

     \begin{scope}[]
       \draw[dashed,<->,red,line width=.66mm] (-1,1)
       to [bend left=70]
       node[midway,right,scale=1.5,align=left] {{\sffamily Block}\\{\sffamily Recycling}}
       (-1,-1.2);
     \end{scope}

     \begin{scope}[shift={(-21,1)}]
       \node[anchor=center,scale=2, align=center, anchor=north] at (4, 1) {Buddy\\Allocator};
       \draw[] (2,2) rectangle (6,-4);
     \end{scope}

     \begin{scope}[shift={(-3,1)}]
       \node[anchor=center,scale=2] at (4, 1) {\mad{}};
       \draw[] (2,2) rectangle (6,-4);
     \end{scope}

   \end{tikzpicture}}








	\caption{Comparison of page allocation for a given sequence with and without \mad{}. Without \mad{}, the second \texttt{alloc} call returns $\psi$, demonstrating enumeration. With \mad{}, block recycling ensures that the second invocation of \texttt{alloc} returns $\phi$ again.}
	\label{fig:page-recycling}
\end{figure}

\section{Background}
\label{s:background}

We expect the reader to be intimately familiar with \rh{} attacks and buddy allocators. Thus, to save space and provide background where needed, we focus on a brief discussion of software diversity.

In \citeyear{cohen}, \citeauthor{cohen} published his pioneering article on software diversity and called it the ``ultimate defense''~\cite{cohen}.
He argued that sufficiently unpredictable execution behavior increases the complexity of attacks such that they are not impossible but become too costly to perform.
In \citeyear{franz+10}, \citeauthor{franz+10} saw that through aligned paradigm shifts, the major obstacles foreseen by \citeauthor{cohen} would be overcome~\cite{franz+10}.
These shifts, along with the advent of code-reuse attacks, led to renewed interest in software diversity~\cite{sok_automated_software_diversity}.
Three different diversification methods have proven capable: (i) virtual-machine based diversification~\cite{williams.etal+09,Hiser2012}, (ii) binary-rewriting based diversity~\cite{Pappas2012}, and (iii) compile-time based diversity~\cite{homescu.etal+13,homescu.etal+15,crane.etal+15a}.
Recently, a hybrid technique combining rewriting and compilers was proposed~\cite{ccr}.

Most of this research, however, focuses exclusively on thwarting arbitrary code execution attacks---with earlier papers focusing on preventing code injection (e.g., \cite{Barrantes2003,Kc2003}) and later ones focusing on preventing code reuse attacks.
A notable exception is \citeauthor{thwarting_side_channels_dynamic_diversity}'s work on using dynamic diversity to prevent timing-based cache side channels~\cite{thwarting_side_channels_dynamic_diversity}.
Also in 2015, \citeauthor{rane.etal+15} presented a way to use principles from obfuscation to close side channels~\cite{rane.etal+15}.

\begin{figure*}[t!]
	\centering
	\scalebox{.6}{%
		\begin{tikzpicture}[>=stealth',shorten >=1pt,auto,node distance=2.8cm,
				semithick]

     \begin{scope}[]
       \draw[-, dashed,red,line width=.66mm] (-4, -.25) rectangle (18.5, 2);
       \node[scale=1.5,align=left,red] at (-5.5, 0.75) {\textsf{Horizontal}\\\textsf{Diversity}};

       \draw[->,dash dot,red,line width=.66mm]
       (9.5,1.3) to [out=150,in=30]
                  node[midway,below,scale=1.5,align=right] {\textsf{A}}
       (7, 1.3);
     \end{scope}

     \begin{scope}[]
       \draw[->, dashed,blue,line width=.66mm] (1.5,-2.5)
       to [out=90,in=-90]
       node[midway,left=.66cm,scale=1.5,align=right] {%
         \begin{tabular}{lr}
           \textsf{Inverse Vertical}  & \multirow{2}{*}{\textsf{C}} \\
           \textsf{Diversity} \\
         \end{tabular}}
       (3.5, 0);
       \draw[->, dashed,blue,line width=.66mm] (1.5,-2.5)
       to [out=90,in=-90]
       node[midway,above,scale=2.0] {}
       (6.5, 0);
     \end{scope}

     \begin{scope}
       \draw[->,dotted,ForestGreen,line width=.66mm] (10,0)
       to [out=-90,in=90]
       node[midway,above,scale=2.0] {}
       (13, -2.5);

       \draw[->,dotted,ForestGreen,line width=.66mm] (14,0)
       to [out=-90,in=90]
       node[midway,right,scale=1.5] {%
         \begin{tabular}{ll}
           \multirow{2}{*}{\textsf{B}} & \textsf{Vertical}  \\
           & \textsf{Diversity} \\
         \end{tabular}}
       (13, -2.5);
     \end{scope}

     \begin{scope}[shift={(0,1.5)}]
       \node[anchor=center,scale=1.5] at (4, 0) {Allocation caches $C_{A,o}$};
       \node[anchor=center,scale=1.5] at (12.5,0) {Shadow caches $C_{S,o}$};
     \end{scope}

     \begin{scope}[shift={(0,0)}]
       \node[anchor=south east,scale=1.5] at (-.5, 0) [align=right]{Order 0,  $C_{A,0}$};
       \node[anchor=south west,scale=1.5] at (17, 0) [align=right]{$C_{S,0}$};
       \draw[] (0,0) rectangle (1, 1) node[midway, anchor=center, align=center] {$b_0$};
       \draw[] (1,0) rectangle (2, 1) ;
       \draw[] (2,0) rectangle (3, 1) node[midway, anchor=center, align=center] {$b_2$};
       \draw[] (3,0) rectangle (4, 1) node[midway, anchor=center, align=center] {\buddyL{p}};
       \draw[] (4,0) rectangle (5, 1) ; 
       \draw[] (5,0) rectangle (6, 1) node[midway, anchor=center, align=center] {$b_5$};
       \draw[] (6,0) rectangle (7, 1) node[midway, anchor=center, align=center] {\buddyR{p}};
       \draw[] (7,0) rectangle (8, 1) node[midway, anchor=center, align=center] {$b_7$};

       \draw[] (8.5,0)  rectangle (9.5, 1);
       \draw[] (9.5,0)  rectangle (10.5, 1) node[midway, anchor=center, align=center] {$b_4$};
       \draw[] (10.5,0) rectangle (11.5, 1) node[midway, anchor=center, align=center] {$b_u$};
       \draw[] (11.5,0) rectangle (12.5, 1) node[midway, anchor=center, align=center] {$b_x$};
       \draw[] (12.5,0) rectangle (13.5, 1);
       \draw[] (13.5,0) rectangle (14.5, 1) node[midway, anchor=center, align=center] {$b_3$};
       \draw[] (14.5,0) rectangle (15.5, 1);
       \draw[] (15.5,0) rectangle (16.5, 1);
     \end{scope}

     \begin{scope}[shift={(0,-3.5)}]
       \node[anchor=south east,scale=1.5] at (-.5, 0) [align=right]{Order 1,  $C_{A,1}$};
       \node[anchor=south west,scale=1.5] at (17, 0) [align=right]{$C_{S,1}$};
       \draw[] (0,0) rectangle (1, 1) node[midway, anchor=center, align=center] {$b'_0$};
       \draw[] (1,0) rectangle (2, 1) node[midway, anchor=center, align=center] {$b'_1$};
       \draw[] (2,0) rectangle (3, 1) node[midway, anchor=center, align=center] {$b'_2$};
       \draw[] (3,0) rectangle (4, 1) node[midway, anchor=center, align=center] {$b'_3$};
       \draw[] (4,0) rectangle (5, 1) node[midway, anchor=center, align=center] {$b'_4$};
       \draw[] (5,0) rectangle (6, 1) node[midway, anchor=center, align=center] {$b'_5$};
       \draw[] (6,0) rectangle (7, 1) node[midway, anchor=center, align=center] {$b'_6$};
       \draw[] (7,0) rectangle (8, 1) node[midway, anchor=center, align=center] {$b'_7$};

       \draw[] (8.5,0)  rectangle (9.5, 1);
       \draw[] (9.5,0)  rectangle (10.5, 1);
       \draw[] (10.5,0) rectangle (11.5, 1);
       \draw[] (11.5,0) rectangle (12.5, 1);
       \draw[] (12.5,0) rectangle (13.5, 1) node[midway, anchor=center, align=center] {$\mathcal{B}$};
       \draw[] (13.5,0) rectangle (14.5, 1);
       \draw[] (14.5,0) rectangle (15.5, 1);
       \draw[] (15.5,0) rectangle (16.5, 1);
     \end{scope}

		\end{tikzpicture}}
	\caption{Block Recycling = Horizontal + Vertical Diversity. %
		Step \textsf{A} shows horizontal diversity, i.e., moving blocks from a shadow cache to the corresponding allocation cache.
		Step \textsf{B} shows vertical diversity, where found buddy blocks $b_{3}$ and $b_{4}$ in shadow cache $C_{S,0}$ will be merged into block $\mathcal{B}$ and put at a random location of the shadow cache of order $C_{S,1}$.
		Step \textsf{C} shows \emph{inverse} vertical diversity, where randomly selected block $b'_1$ of order 1 in $C_{A,1}$ is split up into two blocks of order 0, block \buddyL{x} and block \buddyR{x}, which will be put at random locations in $C_{A,0}$.}
      \label{fig:vdiv}
\end{figure*}
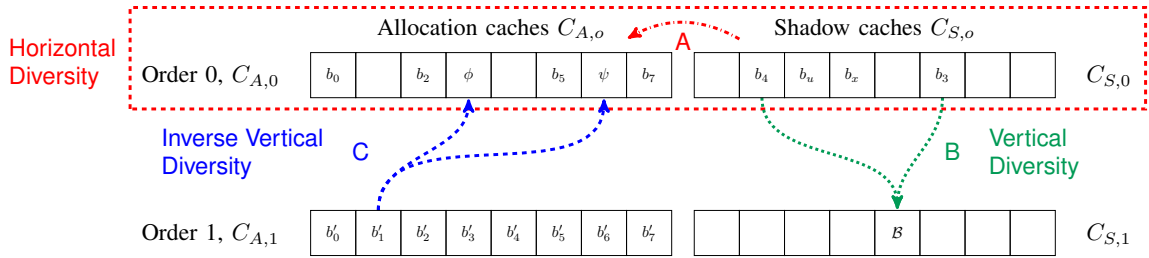

\section{Memory Allocation Diversity}
\label{s:mad}
\mad{} cooperates with existing memory managers, such as the buddy allocator in Linux, but it does not depend on any specific features or properties of a memory manager, i.e., it would also work with a free-list-based memory management system.
Furthermore, \mad{} can work in the operating system's kernel space, i.e., it can handle both memory management requests from the kernel and from user space.

\textbf{Enumerating Memory Blocks}.
A memory manager may provide a functionality that allows to enumerate memory blocks.
Such enumeration is enormously helpful for sparse-allocation massaging, since it ensures that an attacker will be able to process \emph{all} memory eventually.
\autoref{fig:page-recycling} illustrates this behavior on the left-hand side, which shows an allocation sequence and its predictable behavior under memory massaging in a memory manager, a buddy allocator in this case.
Subsequent allocations return different blocks $\phi$ and $\psi$.
On the right-hand side of~\autoref{fig:page-recycling}, however, we see how \mad{} uses so-called \emph{block recycling} to ensure that the second call, with high likelihood, returns the same page $\phi$.
Put differently, to break enumeration and thus delay sparse-allocation massaging, \mad{} increases the block recycling frequency across multiple allocation and free requests.
To this end, \mad{} applies two complementary, mutually-beneficial, spatial diversification techniques, \emph{horizontal} and \emph{vertical diversity}.

\textbf{Horizontal Diversity}.
The goal of block recycling is to ensure that allocation sequences, i.e., a sequence of \texttt{alloc}, \texttt{free}, \texttt{alloc} calls, operate on the same physical blocks to the maximum possible extent.
\mad{} implements a spatial diversification technique that we call \emph{horizontal diversity} (see step \textsf{A} in \autoref{fig:vdiv}), by using two sets of caches:
\begin{itemize}
	\item \emph{allocation caches} ($C_A$), which serve allocation requests;
	\item \emph{shadow caches} ($C_S$), which hold freed blocks.
\end{itemize}

Both allocation and shadow caches exist for all block orders of a buddy allocator, which is shown in~\autoref{fig:vdiv} as a second label in the subscript, e.g., $C_{A,0}$ denotes the allocation cache of order zero, and $C_{S,3}$ denotes the shadow cache of order three.

We use the example from~\autoref{fig:page-recycling} to describe our implementation.
Assume both allocation requests (\texttt{alloc}) as well as the intermediate \texttt{free} request use order zero.
Both allocations will therefore obtain blocks cached in $C_{A,0}$, whereas the \texttt{free} will put the block (page \texttt{p}) into the shadow cache $C_{S,0}$.
When an allocation cache becomes empty, we refill this allocation cache by moving blocks from the corresponding shadow cache of the same order back to the allocation cache---thus \emph{horizontal} diversity.
Both allocations in our example will, therefore, only ever operate on the blocks cached in $C_{A,0}$.
To avoid predictability, allocation and free requests are randomized, i.e., \mad{} fetches a random block from the allocation cache with the proper order.
Conversely, a \texttt{free} request will put the block at a random position in the corresponding shadow cache.

All by itself, horizontal diversity suffers from the following downside.
When the adversary allocates $n + 1$ pages from an allocation cache of size $n$, the allocation request has to be served from the underlying memory manager.
As a result, horizontal diversity by itself would merely delay---but not prevent---enumeration.
A separate technique, \emph{vertical diversity}, is required to address this problem.

\textbf{Vertical Diversity}.
The objectives of vertical diversity are as follows:
(i) provide high utilization to avoid the need for allocating pages from the underlying memory manager,
(ii) provide an alternative, safe way to refill allocation caches,
(iii)~avoid determinism and predictability through randomization.
To achieve these objectives, \mad{} uses the following complementary two techniques.

To maximize cache utilization, \mad{} uses \emph{vertical diversity} (see step \textsf{B} in \autoref{fig:vdiv}), which proactively looks for buddy blocks in a shadow cache of a given order (e.g., $C_{S,0}$ in \autoref{fig:vdiv}).
Found buddies will be merged and put at a random location in the next higher order (e.g., $C_{S,1}$ in \autoref{fig:vdiv}).

To refill an empty allocation cache when the corresponding shadow cache is also empty, \mad{} uses \emph{inverse vertical diversity}.
Step \textsf{C} in \autoref{fig:vdiv} shows an example of inverse vertical diversity in action.
To refill allocation cache $C_{A,0}$, \mad{} randomly selects a block in a higher order ($C_{A,1}$ in \autoref{fig:vdiv}), splits it up into two blocks, and puts those two blocks at random locations in $C_{A,0}$.

Since horizontal diversity moves blocks from a shadow cache to the corresponding allocation cache, the combination of both diversification techniques ensures that benign block allocations and corresponding frees will result in maximum utility and block recycling.

\textbf{Initialization and Refilling}.
\mad{} interacts with the underlying memory manager in the following three situations.
First, \mad{} needs to initialize its own caches when the system becomes active, i.e., during boot or browser startup.
Second, \mad{} needs to be refilled when its caches become empty, as this situation prevents vertical diversity.
To refill its caches, \mad{} obtains pages from the underlying memory manager (e.g., the buddy allocator in Linux).
Third, \mad{} needs to drain its shadow caches when they are full and there is no space left to put freed blocks.
This situation happens, e.g., when a program frees a lot more blocks than can be held in the shadow cache.
To this end, \mad{} returns pages to the underlying memory manager.
If one of these steps is performed in a deterministic fashion, \mad{} would suffer from the penalty of predictability, as the attacker could create an advantageous adversarial configuration to ``feed'' \mad{}.
To address this penalty, \mad{} randomizes all three steps.

\textbf{Diversified Thresholds}.
The exact cache state triggering either horizontal or vertical diversity needs further consideration.
Assume that an attacker knows, e.g., both the lower and upper threshold of elements in the cache are identical and configured as $t$.
The attacker could then create a configuration of allocation caches where the number of elements in each order $C_{A,i}$ is $t+1$ and the corresponding shadow caches $C_{S,i}$ are empty.
By allocating a single block in the lowest order, i.e, order zero, the attacker triggers both horizontal and vertical diversity, in addition to a complete refill of the allocation caches.
\mad{} prevents an attacker from creating such an adversarial configuration by diversifying both lower and upper bound thresholds.

\textbf{Conceptual Detection of \Exhaustive{}-Allocation Massaging}.
\Exhaustive{}-allocation massaging means that the adversary coopts the memory management system to act on its behalf.
If the attacker holds a vulnerable configuration, exhausts all memory, frees the target page, and forces the operating system to allocate sensitive data to the target location, then their privilege escalation attack will succeed.

Besides deterring such attacks, \mad{}'s use of caches creates novel ways of detecting \exhaustive{}-allocation massaging.
Through the lens of the caches, \exhaustive{} memory allocation manifests itself through an increased frequency of asymptomatic \mad{} configurations.
If an attacker exhausts all allocation caches and never frees any pages, then both sets of caches---allocation and shadow caches---will be empty, requiring a refill.
Conversely, if the attacker holds a lot of memory and needs to free it, then the allocation caches will be filled. Once the shadow caches are full, \mad{} will hand all additional memory back to the underlying memory manager.
When compared to just tracking and analyzing what happens in the memory allocator itself, \mad {}'s restriction to a smaller memory area managed through its caches effectively acts as a signal booster.
As a result, a \exhaustive{} memory allocation attack will raise a lot of alarm signals.

A possible response by an attacker could be to interleave malicious \exhaustive{} allocation with periods of ``fake'' benign allocation patterns.
To prevent such a maneuver, \mad{} collects and analyzes multiple information sources.
First, \mad{} collects and measures the frequency of occurring asymptomatic configurations.
Second, \mad{} uses diversified snapshot intervals, i.e., it analyzes its own caches every $n$ allocations, where $n$ is a randomized interval.
Since this snapshot collection can be efficiently implemented, we can configure the random snapshot interval to be low.
In our experiments, for example, we use a random number in the range $[13, 997]$.
A prototypical implementation of this technique detects virtually all \exhaustive{}-allocation massaging (see column ``Detection Rate'' in~\autoref{tab:worst-case-allocs-vuln-cfg}).

The outlined detection technique combines (i) high-resolution monitoring of memory allocation activity with (b) principles of software diversity to counter evasion attempts.

\textbf{Generalization}
Although our discussion so far relied primarily on cooperation with a buddy allocator in an operating system, the core principles of \mad{} generalize to other applications and memory managers.
\mad{} sits on top of a memory allocator and, therefore, does not require a buddy allocator, but could also be combined with a much simpler free-list memory-allocator.
Furthermore, \mad{} is not tied to any specific operating system internals and can be used in any application that performs its own memory management, such as web browsers, database systems, or virtual machines.
\mad{} segments memory into blocks of different order, but the specific geometry can be tailored to an application's specific use case.
In a web browser, for example, \mad{} could be used to manage the JavaScript heap, thus deterring JavaScript-based \rh{} attacks.


\section{Evaluation}
\label{s:eval}

\subsection{Quantitative Security}
\label{ss:eval-security}

This section details the results of the security experiments to evaluate \mad{}.
Specifically, we want to evaluate \mad{}'s efficiency at preventing sparse-allocation massaging.

\begin{figure}[t!]
	\centering
	\includegraphics[width=\linewidth]{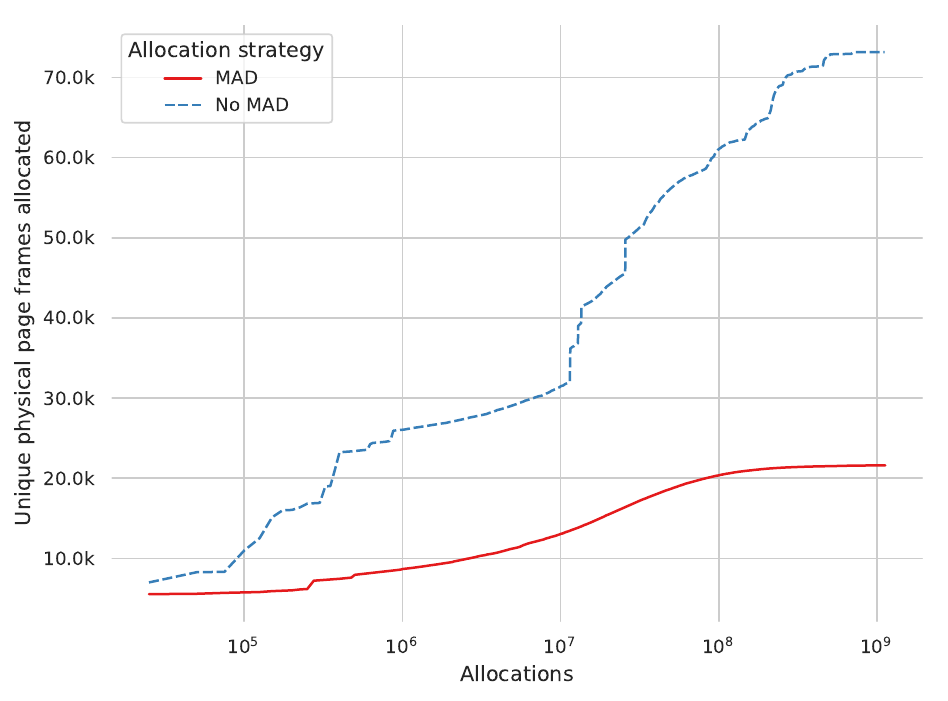}
	\caption{Comparison of \mad{} with a textbook buddy allocator under sparse-allocation massaging.}
	\label{fig:mad-cmp-sparse}
\end{figure}

We evaluated the efficiency of \mad{} by massaging memory in a randomized fashion, inspired by memory chasing.
We measured the amount of unique physical memory blocks obtained when running one billion memory allocations against both a textbook buddy allocator and \mad{}, at intervals of 25,000 allocations.
The results of this experiment are shown in~\autoref{fig:mad-cmp-sparse}.
Note the salient point of \mad{} showing a plateau of unique physical blocks allocated between 100 million and 1 billion allocations.
Put differently, for over 900 million allocations, \mad{} did not yield substantially more blocks.

To quantify the attrition rate of unique physical blocks per number of allocations, we computed the difference in number of unique physical blocks allocated per 100,000 allocations over the 1~billion allocations measured.
On average, \mad{}'s attrition rate is 0.3563 unique physical blocks per 25,000 allocations.
Our baseline buddy allocator's attrition rate is 1.4832 unique physical blocks per 25,000 allocations.
\mad{} improves the attrition rate by a factor of 4.16$\times$.

Extrapolated on the 4 million physical memory blocks present in a system with 16 GB of memory, complete enumeration of all memory blocks without \mad{} would require, on average, about 77 billion allocations.
Using \mad{}, this number of allocations increases by an order of magnitude to an average of 294 billion allocations.

\begin{figure}[t!]
	\centering
	\includegraphics[width=\linewidth]{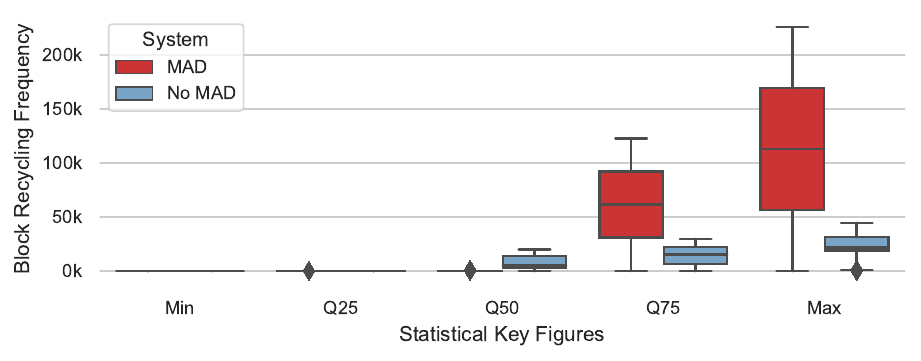}
	\caption{Comparison of block recycling frequency of \mad{} with a textbook buddy allocator.%
    }
	\label{fig:mad-cmp-prf}
\end{figure}

\autoref{fig:mad-cmp-prf} shows the different block recycling frequencies of MAD vs.~our buddy allocator.
\autoref{fig:mad-cmp-sparse} indicates that after a billion allocations we have merely allocated about 70,000 of a total of about 4 million blocks.
\autoref{fig:mad-cmp-prf} shows that the majority of blocks in minimum, first and second quartile is zero, meaning that most memory blocks of the system have not been allocated at all.
The significant increase in block recycling frequencies manifests itself in the third quartile and the maximum, meaning that some memory blocks were allocated much more frequently than others.
We find that the maximum block recycling frequencies between \mad{} and the buddy allocator differ by a factor of four.

\subsection{Probability of Success Under Worst Case Assumptions}
A worst case assumption for \mad{} is when an attacker succeeds to get control over the blocks required to perform \rh{}, and manages to let \mad{} allocate the memory at the target location.
Due to randomizing initialization and refilling (see~\autoref{s:mad}), we were not able to put a specific block into \mad{}'s allocation caches and, therefore, had to resort to choose a random block of order six and designate it as a vulnerable page.
Recall that \mad{} is hardware agnostic, and does not actually carry any specific information about physical-to-DRAM address mapping.
Without loss of generality, we assume that a vulnerable block configuration, i.e., a vulnerable block and its logical neighbors, is determined solely by their physical block numbers.

We evaluated \mad{}'s deterrence by measuring the number of allocations required to obtain a vulnerable configuration.
\autoref{tab:worst-case-allocs-vuln-cfg} contains our results, including averages and medians over 50 repetitions.
Note that the average and median number of required allocations differ substantially.
This difference is due to the probabilistic nature of \mad{}:
sometimes pages required for an attack get moved to a shadow cache or recycled to higher orders via vertical diversity.

The probability of success varies with the lower and upper bounds used to allocate memory blocks and ranges between less than 0.3\% and less than 0.01\%.
As a result, more than 99.5\% of the times, an adversary will not succeed to predictably force allocation into the target location.

\begin{table}[t!]
  \centering
  \begin{tabular}{@{}rrcrrr@{}}
    \toprule
    \multicolumn{2}{c}{Alloc. Size} & & \multicolumn{2}{c}{Required Alloc.} & Detection\\
    \cmidrule{1-2}
        \cmidrule{4-5}
    \multicolumn{1}{c}{LB} & \multicolumn{1}{c}{UB} & & \multicolumn{1}{c}{Average} & \multicolumn{1}{c}{Median} & \multicolumn{1}{c}{Rate}\\
    \midrule
    4  & 8   & & 597,301 & 112,916 &  98\% \\
    8  & 16  & & 565,959 & 70,065  &  100\% \\
    16 & 32  & & 418,565 & 37,858  &  100\% \\
    32 & 64  & & 278,490 & 30,729  &  100\% \\
    64 & 128 & & 255,154 & 39,563  &  98\% \\
    \bottomrule
  \end{tabular}
  \caption{Number of required allocations to obtain a vulnerable configuration assuming the attacker succeeded in placing a page into \mad{}.
    LB and UB indicate lower and upper bounds used for memory allocation size.}
  \label{tab:worst-case-allocs-vuln-cfg}
\end{table}

\subsection{Implementation Complexity}
\label{ss:eval-simplicity}
We have implemented \mad{} in Python (i) to guide the search for optimal parameters that, e.g.,  determine the sizes of allocation and shadow caches, and (ii) to create heat-map videos of the overall memory, and videos showing the internal state of \mad{}'s allocation caches.
The resulting Python implementation uses roughly a thousand lines of code.

\section{Related Work}
\label{s:related-work}
Since \citeauthor{rowhammer}~discovered the \rh{} vulnerability in \citeyear{rowhammer}, research communities in various fields published a plethora of papers on \rh{}.
\citeauthor{mutlu2019rowhammer}~provide an extensive review over the state-of-the-art~\cite{mutlu2019rowhammer}.

Prior work has proposed several defenses to protect systems against \rh{} attacks without requiring hardware replacement.
These defenses either only protect against attacks targeting specific memory areas or critically depend on knowledge about the deployed DIMMs.
None of these defenses, furthermore, prevent many-sided \rh{} or would require prohibitive amount of memory to do so.

\subsubsection{Kernel and User Space Separation}
\label{ss:kernel_and_userspace_seperation}

G-CATT~\cite{gcatt} physically separates memory in at least two areas: a kernel area and a user-space area.
Conceptually, this approach prevents attacks targeting kernel space, but it cannot protect against attacks that induce bit flips in user space.
\citeauthor{another_flip_in_the_wall} presented \emph{opcode flipping}, an attack exploiting bit flips in user space mapped binaries, such as \texttt{sudo}~\cite{another_flip_in_the_wall}.
Furthermore, \citeauthor{still_hammerable}~showed that G-CATT cannot fully protect against attacks targeting double-owned kernel buffers~\cite{still_hammerable}.

\subsubsection{Defenses Based on Memory Layout}
\label{ss:memory_layout_based_defenses}
ZebRAM~\cite{zebram} and ALIS~\cite{throwhammer} use \emph{unsafe regions} and \emph{guard rows}, respectively.
For that purpose, they require knowledge about the physical-to-DRAM address mapping, which is generally not published by the vendors---some reverse engineered documentation is available~\cite{drama, throwhammer, anvil, crossVM_rh, Wang2020}.

GuardION~\cite{rampage_guardion} effectively prevents \rh{} attacks on Android devices by inserting so-called \emph{guard pages} before and after contiguous direct memory access (DMA) memory regions.
However, on other popular architectures (such as x86) attackers can row hammer without direct uncached memory access~\cite{rowhammer,anvil,project_zero_rowhammer,rowhammerjs,another_flip_in_the_wall,Frigo2020,Ridder2021}.

\citeauthor{monotonic_pointers} use monotonic pointers~\cite{monotonic_pointers} to protect page tables from \rh{} attacks and thus cannot protect other targets.
In addition, defenders need to know vulnerable memory cells and their flip direction in advance.

\subsubsection{Defenses Preventing Bit Flips}
\label{sec:defenses-prev-bit}

ANVIL~\cite{anvil} effectively prevents \rh{} bit flips triggered by software by refreshing physically neighboring memory cells of a potential victim cell if row hammering is detected.
ANVIL relies on the Intel Performance Counter Monitor~\cite{intel_perf_monitor}, which is not accurate enough for security critical applications~\cite{performance_counters_non_determinism}, as well as knowledge about the inner, hard-wired chip design of the memory modules.

B-CATT~\cite{bcatt} prevents the operating system from using blocks vulnerable to row hammering.
B-CATT blacklists vulnerable blocks during boot time, which can thus increase significantly---attackers can still use bit flips undetected by B-CATT.
Since many memory modules have one bit flip per page, the entire memory has to be blacklisted~\cite{drammer, crossVM_rh, another_flip_in_the_wall}.

\section{Conclusions \& Future Work}
No known defense mitigates the more recently discovered many-sided \rh{} attacks.
Based on previous years and ongoing developments, it is unlikely that we, as a community, know all the facets and relevant details of how \rh{} and, more generally, DRAM attacks will evolve.
In light of these developments, early defenses that focused on identifying and/or isolating aggressor from victim rows fail to generalize from double-sided to many-sided \rh{}.
The principle of isolation is doomed, as it would require much more memory to scale from double-sided to many-sided \rh{} attacks, resulting in prohibitive expensive memory requirements.

\mad{} presents a new research direction that brings ideas underlying software diversity to the idea of mitigating \rh{} attacks in memory management components.
As is true for all other defenses using software diversity, \mad{} offers probabilistic security, i.e., a brute-force attack will succeed eventually.
To protract the time required for such a brute-force attack, \mad{} combines horizontal and vertical diversity, and our preliminary data provides promising evidence of \mad{}'s protraction capabilities.
Besides protracting attacks, an operating system using \mad{} may also be able to leverage the uncertainty introduced to detect attacks.
If, for example, a \rh{} attack intends to escalate privileges and \mad{} does not put the expected data into the target location, the attack will manipulate other data and a subsequent access requiring higher privileges will fail.

\mad{} does not require specific hard- or software information to operate.
This conceptual simplicity is also beneficial when going from one- or double-sided to many-sided \rh{} attacks:
From the perspective of \mad{}, it does not matter how many rows an attacker needs, and since a many-sided \rh{} attack requires control of \emph{more rows}, \mad{}'s deterrence may actually also be \emph{more effective}.

Based on the encouraging evidence, a more thorough investigation of \mad{} is warranted.
We plan on implementing \mad{} in an operating system and a web browser, and subsequently evaluate \mad{}'s protective properties against all known \rh{} attacks.
We also believe that a hybrid technique that combines \mad{} with another, stronger but more expensive defense holds potential to mitigate attacks.
Besides examining real-world attacks, we plan on investigating a variety of properties of \mad{}'s probabilistic caches, such as fragmentation, detection, and steady cache states.


\section*{Acknowledgments}
This paper has been partly supported by two Austrian Federal Ministries (BMK and BMDW), and the Province of Upper Austria in the frame of the COMET center SCCH, grant no. FFG-865891.
Parts of the research have been financed by research subsidies granted by the Province of Upper Austria in the project \emph{DEPS Pilot}.
This project has received funding from the European Union's Horizon 2020 research and innovation programme under grant agreement No 830927.

\balance
\printbibliography

@inproceedings{Ridder2021,
  author    = {Finn de Ridder and Pietro Frigo and Emanuele Vannacci and Herbert Bos and Cristiano Giuffrida and Kaveh Razavi},
  title     = {{SMASH: Synchronized Many-sided Rowhammer Attacks from JavaScript}},
  booktitle = {30th {USENIX} Security Symposium ({USENIX} Security)},
  year      = {2021}
}

@inproceedings{Frigo2020,
  author    = {Frigo, Pietro and Vannacci, Emanuele and Hassan, Hasan and van der Veen, Victor and Mutlu, Onur and Giuffrida, Cristiano and Bos, Herbert and Razavi, Kaveh},
  title     = {{TRRespass: Exploiting the Many Sides of Target Row Refresh}},
  booktitle = {IEEE Symposium on Security and Privacy (S\&P)},
  year      = {2020}
}

@InProceedings{Wang2020,
  author     = {Wang, Minghua and Zhang, Zhi and Cheng, Yueqiang and Nepal, Surya},
  booktitle  = {Proceedings of the 57th ACM/EDAC/IEEE Design Automation Conference (DAC)},
  title      = {{DRAMDig: A Knowledge-assisted Tool to Uncover DRAM Address Mapping}},
  date       = {2020},
}

@inproceedings{rowhammerjs,
  author    = {Gruss, Daniel and Maurice, Cl{\'e}mentine and Mangard, Stefan},
  title     = {{Rowhammer.js: A Remote Software-Induced Fault Attack in JavaScript}},
  booktitle = {Proceedings of the 13th International Conference on Detection of Intrusions and Malware, and Vulnerability Assessment (DIMVA)},
  year      = {2016}
}

@inproceedings{drammer,
  author    = {Victor van der Veen and Yanick Fratantonio and Martina Lindorfer and Daniel Gruss and Cl\'ementine Maurice and Giovanni Vigna and Herbert Bos and Kaveh Razavi and Cristiano Giuffrida},
  title     = {{Drammer: Deterministic Rowhammer Attacks on Mobile Platform}},
  booktitle = {Proceedings of the 23rd Conference on Computer and Communications Security (CCS)},
  year      = {2016},
  month     = {10}
}

@inproceedings{gcatt,
  author    = {Brasser, Ferdinand and Davi, Lucas and Gens, David and Liebchen, Christopher and Sadeghi, Ahmad-Reza},
  title     = {{CAn'T Touch This: Software-only Mitigation Against Rowhammer Attacks Targeting Kernel Memory}},
  booktitle = {Proceedings of the 26th USENIX Conference on Security Symposium (USENIX Security)},
  year      = {2017}
}

@article{bcatt,
  author        = {Ferdinand Brasser and
               Lucas Davi and
               David Gens and
               Christopher Liebchen and
               Ahmad{-}Reza Sadeghi},
  title         = {{CAn't Touch This: Practical and Generic Software-only Defenses Against
               Rowhammer Attacks}},
  journal       = {CoRR},
  year          = {2016},
  archiveprefix = {arXiv},
  eprint        = {1611.08396}
}

@inproceedings{anvil,
  author    = {Aweke, Zelalem Birhanu and Yitbarek, Salessawi Ferede and Qiao, Rui and Das, Reetuparna and Hicks, Matthew and Oren, Yossi and Austin, Todd},
  title     = {{ANVIL: Software-Based Protection Against Next-Generation Rowhammer Attacks}},
  booktitle = {Proceedings of the Twenty-First International Conference on Architectural Support for Programming Languages and Operating Systems (ASPLOS)},
  year      = {2016}
}

@online{project_zero_rowhammer,
  url    = {https://googleprojectzero.blogspot.com/2015/03/exploiting-dram-rowhammer-bug-to-gain.html},
  year   = {2015},
  month  = {3},
  title  = {{Exploiting the DRAM rowhammer bug to gain kernel privileges}},
  author = {Mark Seaborn and Thomas Dullien}
}

@inproceedings{another_flip_in_the_wall,
  author    = {Daniel Gruss and
               Moritz Lipp and
               Michael Schwarz and
               Daniel Genkin and
               Jonas Juffinger and
               Sioli O'Connell and
               Wolfgang Schoechl and
               Yuval Yarom},
  title     = {{Another Flip in the Wall of Rowhammer Defenses}},
  booktitle = {IEEE Symposium on Security and Privacy (S\&P)},
  year      = {2018}
}

@inproceedings{rampage_guardion,
  author    = {Victor van der Veen and Martina Lindorfer and Yanick Fratantonio and Harikrishnan Padmanabha Pillai and Giovanni Vigna and Christopher Kruegel and Herbert Bos and Kaveh Razavi},
  title     = {{GuardION: Practical Mitigation of DMA-based Rowhammer Attacks on ARM}},
  booktitle = {Proceedings of the 15th Conference on Detection of Intrusions and Malware, and Vulnerability Assessment (DIMVA)},
  year      = {2018},
  month     = {6}
}

@inproceedings{rowhammer,
  author    = {Kim, Yoongu and Daly, Ross and Kim, Jeremie and Fallin, Chris and Lee, Ji Hye and Lee, Donghyuk and Wilkerson, Chris and Lai, Konrad and Mutlu, Onur},
  title     = {{Flipping Bits in Memory Without Accessing Them: An Experimental Study of DRAM Disturbance Errors}},
  booktitle = {Proceeding of the 41st Annual International Symposium on Computer Architecuture (ISCA)},
  year      = {2014}
}

@inproceedings{flip_feng_shui,
  author    = {Kaveh Razavi and Ben Gras and Erik Bosman and Bart Preneel and Cristiano Giuffrida and Herbert Bos},
  title     = {{Flip Feng Shui: Hammering a Needle in the Software Stack}},
  booktitle = {25th USENIX Security Symposium (USENIX Security)},
  year      = {2016}
}

@inproceedings{throwhammer,
  author    = {Andrei Tatar and Radhesh Krishnan Konoth and Elias Athanasopoulos and Cristiano Giuffrida and Herbert Bos and Kaveh Razavi},
  title     = {{Throwhammer: Rowhammer Attacks over the Network and Defenses}},
  booktitle = {{USENIX} Annual Technical Conference ({USENIX} {ATC})},
  year      = {2018}
}

@inproceedings{sok_automated_software_diversity,
  author    = {Per Larsen and
               Andrei Homescu and
               Stefan Brunthaler and
               Michael Franz},
  title     = {{SoK: Automated Software Diversity}},
  booktitle = {IEEE Symposium on Security and Privacy (S\&P)},
  year      = {2014}
}

@article{cohen,
  author  = {Frederick Cohen},
  title   = {{Operating System Protection Through Program Evolution}},
  journal = {Computers and Security},
  month   = {10},
  year    = {1993}
}

@inproceedings{thwarting_side_channels_dynamic_diversity,
  title     = {{Thwarting Cache Side-Channel Attacks Through Dynamic Software Diversity.}},
  author    = {Stephen Crane and Andrei Homescu and Stefan Brunthaler and Per Larsen and Michael Franz},
  booktitle = {22nd Annual Network and Distributed System Security Symposium (NDSS)},
  year      = {2015}
}

@online{intel_perf_monitor,
  title  = {{Intel® Performance Counter Monitor - A Better Way to Measure CPU Utilization}},
  url    = {https://software.intel.com/en-us/articles/intel-performance-counter-monitor},
  author = {Thomas Willhalm and Roman Dementiev and Patrick Fay},
  month  = {1},
  year   = {2017}
}

@article{still_hammerable,
  author        = {Yueqiang Cheng and Zhi Zhang and Surya Nepal},
  title         = {{Still Hammerable and Exploitable: on the Effectiveness of Software-only Physical Kernel Isolation}},
  journal       = {CoRR},
  year          = {2018},
  archiveprefix = {arXiv},
  eprint        = {1802.07060}
}

@inproceedings{zebram,
  author    = {Radhesh Krishnan Konoth and Marco Oliverio and Andrei Tatar and Dennis Andriesse and Herbert Bos and Cristiano Giuffrida and Kaveh Razavi},
  title     = {{ZebRAM: Comprehensive and Compatible Software Protection Against Rowhammer Attacks}},
  booktitle = {13th {USENIX} Symposium on Operating Systems Design and Implementation ({OSDI})},
  year      = {2018}
}

@inproceedings{ccr,
  author    = {Hyungjoon Koo and Yaohui Chen and Long Lu and Vasileios~P. Kemerlis and Michalis Polychronakis},
  title     = {{Compiler-assisted Code Randomization}},
  booktitle = {{IEEE} Symposium on Security and Privacy (S\&P)},
  _pages    = {472--488},
  month     = {May},
  year      = {2018},
  _location = {San Francisco, CA}
}

@inproceedings{monotonic_pointers,
  author    = {Wu, Xin-Chuan and Sherwood, Timothy and Chong, Frederic T. and Li, Yanjing},
  title     = {{Protecting Page Tables from RowHammer Attacks Using Monotonic Pointers in DRAM True-Cells}},
  booktitle = {Proceedings of the Twenty-Fourth International Conference on Architectural Support for Programming Languages and Operating Systems ({ASPLOS})},
  year      = {2019}
}

@inproceedings{crossVM_rh,
  author    = {Yuan Xiao and Xiaokuan Zhang and Yinqian Zhang and Radu Teodorescu},
  title     = {{One Bit Flips, One Cloud Flops: Cross-VM Row Hammer Attacks and Privilege Escalation}},
  booktitle = {25th {USENIX} Security Symposium ({USENIX} Security)},
  year      = {2016}
}

@inproceedings{eccploit,
  title     = {{Exploiting Correcting Codes: On the Effectiveness of ECC Memory Against Rowhammer Attacks }},
  author    = {Cojocar, Lucian and Razavi, Kaveh and Giuffrida, Cristiano and Bos, Herbert},
  booktitle = {IEEE Symposium on Security and Privacy (S\&P)},
  year      = {2019}
}

@inproceedings{performance_counters_non_determinism,
  author    = {Sanjeev Das and J. Werner and M. Antonakakis and M. Polychronakis and F. Monrose},
  booktitle = {IEEE Symposium on Security and Privacy (S\&P)},
  title     = {{SoK: The Challenges, Pitfalls, and Perils of Using Hardware Performance Counters for Security}},
  year      = {2019}
}

@inproceedings{rane.etal+15,
  author    = {Ashay Rane and
               Calvin Lin and
               Mohit Tiwari},
  title     = {{Raccoon: Closing Digital Side-Channels through Obfuscated Execution}},
  booktitle = {24th {USENIX} Security Symposium ({USENIX} Security)},
  _pages    = {431--446},
  year      = {2015}
}

@article{mutlu2019rowhammer,
  title   = {{RowHammer: A Retrospective}},
  author  = {Mutlu, Onur and Kim, Jeremie S},
  journal = {IEEE Transactions on Computer-Aided Design of Integrated Circuits and Systems ({TCAD})},
  year    = {2019}
}

@inproceedings{drama,
  author    = {Peter Pessl and Daniel Gruss and Cl{\'e}mentine Maurice and Michael Schwarz and Stefan Mangard},
  title     = {{DRAMA: Exploiting DRAM Addressing for Cross-CPU Attacks}},
  booktitle = {25th {USENIX} Security Symposium ({USENIX} Security)},
  year      = {2016}
}

@article{homescu.etal+15,
author = {Homescu, Andrei and Jackson, Todd and Crane, Stephen and Brunthaler, Stefan and Larsen, Per and Franz, Michael},
doi = {10.1109/TDSC.2015.2433252},
issn = {1545-5971},
journal = {IEEE Transactions on Dependable and Secure Computing},
keywords = {sbrLeadAuthor},
month = {jun},
number = {2},
_pages = {1--1},
title = {{Large-scale Automated Software Diversity--Program Evolution Redux}},
url = {http://ieeexplore.ieee.org/lpdocs/epic03/wrapper.htm?arnumber=7122891},
volume = {14},
year = {2017}
}

@inproceedings{franz+10,
_address = {New York, New York, USA},
author = {Franz, Michael},
booktitle = {Proceedings of the 2010 New Security Paradigms Workshop (NSPW '10)},
doi = {10.1145/1900546.1900550},
isbn = {9781450304153},
_pages = {7},
_publisher = {ACM Press},
title = {{E unibus pluram}},
url = {http://portal.acm.org/citation.cfm?doid=1900546.1900550},
year = {2010}
}

@inproceedings{crane.etal+15a,
author = {Crane, Stephen and Liebchen, Christopher and Homescu, Andrei and Davi, Lucas and Larsen, Per and Sadeghi, Ahmad-Reza and Brunthaler, Stefan and Franz, Michael},
booktitle = {IEEE Symposium on Security and Privacy (S\&P)},
doi = {10.1109/SP.2015.52},
isbn = {978-1-4673-6949-7},
keywords = {JIT-ROP,defenses,diversity,security},
mendeley-tags = {JIT-ROP,defenses,diversity,security},
month = {may},
_pages = {763--780},
_publisher = {IEEE},
title = {{Readactor: Practical Code Randomization Resilient to Memory Disclosure}},
url = {http://ieeexplore.ieee.org/lpdocs/epic03/wrapper.htm?arnumber=7163059},
year = {2015}
}

@inproceedings{Kc2003,
abstract = {We describe a new, general approach for safeguarding systems against any type of code-injection attack. We apply Kerckhoff's principle, by creating process-specific randomized instruction sets (e.g., machine instructions) of the system executing potentially vulnerable software. An attacker who does not know the key to the randomization algorithm will inject code that is invalid for that randomized processor, causing a runtime exception. To determine the difficulty of integrating support for the proposed mechanism in the operating system, we modified the Linux kernel, the GNU binutils tools, and the bochs-x86 emulator. Although the performance penalty is significant, our prototype demonstrates the feasibility of the approach, and should be directly usable on a suitable-modified processor (e.g., the Transmeta Crusoe). Our approach is equally applicable against code-injecting attacks in scripting and interpreted languages, e.g., web-based SQL injection. We demonstrate this by modifying the Perl interpreter to permit randomized script execution. The performance penalty in this case is minimal. Where our proposed approach is feasible (i.e., in an emulated environment, in the presence of programmable or specialized hardware, or in interpreted languages), it can serve as a low-overhead protection mechanism, and can easily complement other mechanisms. Copyright 2003 ACM.},
author = {Kc, Gaurav S. and Keromytis, Angelos D. and Prevelakis, Vassilis},
booktitle = {Proceedings of the 10th Conference on Computer and Communications Security (CCS)},
doi = {10.1145/948143.948146},
isbn = {1581137389},
issn = {15437221},
keywords = {buffer overflows,emulators,interpreters},
_pages = {272----280},
title = {{Countering Code-Injection Attacks With Instruction-Set Randomization}},
url = {http://portal.acm.org/citation.cfm?doid=948109.948146},
year = {2003}
}

@inproceedings{homescu.etal+13,
author = {Homescu, Andrei and Neisius, Steven and Larsen, Per and Brunthaler, Stefan and Franz, Michael},
booktitle = {Proceedings of the 2013 IEEE/ACM International Symposium on Code Generation and Optimization (CGO)},
doi = {10.1109/CGO.2013.6494997},
isbn = {978-1-4673-5525-4},
keywords = {sbrLeadAuthor,selected},
month = {feb},
_pages = {1--11},
_publisher = {IEEE},
title = {{Profile-guided automated software diversity}},
url = {http://ieeexplore.ieee.org/lpdocs/epic03/wrapper.htm?arnumber=6494997},
year = {2013}
}

@inproceedings{Pappas2012,
abstract = {The wide adoption of non-executable page protec- tions in recent versions of popular operating systems has given rise to attacks that employ return-oriented programming (ROP) to achieve arbitrary code execution without the injection of any code. Existing defenses against ROP exploits either require source code or symbolic debugging information, or impose a significant runtime overhead, which limits their applicability for the protection of third-party applications. In this paper we present in-place code randomization, a practical mitigation technique against ROP attacks that can be applied directly on third-party software. Our method uses various narrow-scope code transformations that can be applied statically, without changing the location of basic blocks, allowing the safe randomization of stripped binaries even with partial disassembly coverage. These transformations effectively eliminate about 10%, and probabilistically break about 80% of the useful instruction sequences found in a large set of PE files. Since no additional code is inserted, in-place code randomization does not incur any measurable runtime overhead, enabling it to be easily used in tandem with existing exploit mitigations such as address space layout randomization. Our evaluation using publicly available ROP exploits and two ROP code generation toolkits demonstrates that our technique prevents the exploitation of the tested vulnerableWindows 7 applications, including Adobe Reader, as well as the automated construction of alternative ROP payloads that aim to circumvent in-place code randomization using solely any remaining unaffected instruction sequences.},
author = {Pappas, Vasilis and Polychronakis, Michalis and Keromytis, Angelos D.},
booktitle = {IEEE Symposium on Security and Privacy (S\&P)},
doi = {10.1109/SP.2012.41},
isbn = {978-1-4673-1244-8},
issn = {10816011},
month = {may},
_pages = {601--615},
_publisher = {IEEE},
title = {{Smashing the Gadgets: Hindering Return-Oriented Programming Using In-place Code Randomization}},
url = {http://ieeexplore.ieee.org/lpdocs/epic03/wrapper.htm?arnumber=6234439},
year = {2012}
}

@inproceedings{Barrantes2003,
abstract = {Binary code injection into an executing program is a common form of attack. Most current defenses against this form of attack use a ‘guard all doors' strategy, trying to block the avenues by which ex- ecution can be diverted. We describe a complementary method of protection, which disrupts foreign code execution regardless of how the code is injected. A unique and private machine instruction set for each executing program would make it difﬁcult for an outsider to design binary attack code against that program and impossible to use the same binary attack code against multiple machines. As a proof of concept, we describe a randomized instruction set em- ulator (RISE), based on the open-source Valgrind x86-to-x86 bi- nary translator. The prototype disrupts binary code injection attacks against a program without requiring its recompilation, linking, or access to source code. The paper describes the RISE implemen- tation and its limitations, gives evidence demonstrating that RISE defeats common attacks, considers how the dense x86 instruction set affects the method, and discusses potential extensions of the idea.},
_address = {New York, New York, USA},
author = {Barrantes, Elena Gabriela and Ackley, David H. and Palmer, Trek S. and Stefanovic, Darko and Zovi, Dino Dai},
booktitle = {Proceedings of the 10th ACM Conference on Computer and Communication Security (CCS)},
doi = {10.1145/948109.948147},
isbn = {1581137389},
issn = {15437221},
keywords = {automated diversity,emulation,information hiding,language randomization,obfuscation,security},
_pages = {281},
_publisher = {ACM Press},
title = {{Randomized Instruction Set Emulation to Disrupt Binary Code Injection Attacks}},
url = {http://dl.acm.org/citation.cfm?id=948109.948147 http://portal.acm.org/citation.cfm?doid=948109.948147},
year = {2003}
}

@article{williams.etal+09,
author = {Williams, Daniel and Hu, Wei and Davidson, Jack W. and Hiser, Jason D. and Knight, John C. and Nguyen-Tuong, Anh},
doi = {10.1109/MSP.2009.18},
issn = {1540-7993},
journal = {IEEE Security \& Privacy Magazine},
month = {jan},
number = {1},
_pages = {26--33},
title = {{Security through Diversity: Leveraging Virtual Machine Technology}},
url = {http://ieeexplore.ieee.org/lpdocs/epic03/wrapper.htm?arnumber=4768651},
volume = {7},
year = {2009}
}

@inproceedings{Hiser2012,
abstract = {Through randomization of the memory space and the confinement of code to non-data pages, computer security researchers have made a wide range of attacks against program binaries more difficult. However, attacks have evolved to exploit weaknesses in these defenses. To thwart these attacks, we introduce a novel technique called Instruction Location Randomization (ILR). Conceptually, ILR randomizes the location of every instruction in a program, thwarting an attacker's ability to re-use program functionality (e.g., arc-injection attacks and return-oriented programming attacks). ILR operates on arbitrary executable programs, requires no compiler support, and requires no user interaction. Thus, it can be automatically applied post-deployment, allowing easy and frequent re-randomization. Our preliminary prototype, working on 32-bit x86 Linux ELF binaries, provides a high degree of entropy. Individual instructions are randomly placed within a 31-bit address space. Thus, attacks that rely on a priori knowledge of the location of code or derandomization are not feasible. We demonstrated ILR's defensive capabilities by defeating attacks against programs with vulnerabilities, including Adobe's PDF viewer, acroread, which had an in-the-wild vulnerability. Additionally, using an industry-standard CPU performance benchmark suite, we compared the run time of prototype ILR-protected executables to that of native executables. The average run-time overhead of ILR was 13% with more than half the programs having effectively no overhead (15 out of 29), indicating that ILR is a realistic and cost-effective mitigation technique.},
author = {Hiser, Jason and Nguyen-Tuong, Anh and Co, Michele and Hall, Matthew and Davidson, Jack W.},
booktitle = {IEEE Symposium on Security and Privacy (S\&P)},
doi = {10.1109/SP.2012.39},
isbn = {978-1-4673-1244-8},
issn = {10816011},
keywords = {ASLR,Diversity,Exploit prevention,Randomization,Return-oriented-programming,arc-injection},
month = {may},
_pages = {571--585},
_publisher = {IEEE},
title = {{ILR: Where'd My Gadgets Go?}},
url = {http://ieeexplore.ieee.org/lpdocs/epic03/wrapper.htm?arnumber=6234437},
year = {2012}
}

\end{document}